\documentstyle[sprocl,epsf]{article}

\def\be{\begin{equation}}
\def\ee{\end{equation}}
\def\bea{\begin{eqnarray}}
\def\eea{\end{eqnarray}}
\def\simm#1{\mathop{\vtop{\ialign{##\crcr
        $\hfil\displaystyle{#1}\hfil$\crcr\noalign{\kern0.5pt\nointerlineskip}
        $\sim$\crcr\noalign{\kern0.5pt}}}}\limits}

\begin{document}

\title{PHASE STRUCTURE OF LATTICE QCD\\
 FOR GENERAL NUMBER OF FLAVORS}

\author{ Y. IWASAKI }

\address{Center for Computational Physics and Institute of Physics\\
University of Tsukuba,Ibaraki 305, Japan}


\maketitle\abstracts{
We investigate the phase structure of lattice QCD for
general number of flavors $N_F$.
Based on numerical results combined with the result of the
perturbation theory
we propose the following picture:
When $N_F \ge 17$, there is only
a trivial fixed point and therefore the theory in the continuum limit
is trivial. On the other hand, when $16 \ge N_F \ge 7$,
there is a non-trivial fixed point and therefore the theory is
non-trivial with anomalous dimensions, however, without quark confinement.
Theories which satisfy both quark confinement and spontaneous
chiral symmetry breaking in the continuum limit
exist only for $N_F \le 6$.}
  

\section{Introduction}
The fundamental properties of QCD are quark confinement,
asymptotic freedom and spontaneous breakdown of chiral symmetry.
It is well-known that if the number of flavors exceeds  $16 \frac{1}{2}$,
the asymptotic freedom is lost.
Then, the question which naturally arises is whether
there is constraint on the number of flavors for
quark confinement and/or the spontaneous breakdown of chiral symmetry.

In our previous work \cite{previo}, 
it was shown that in the strong
coupling limit, when the number of flavors $N_F$ is greater than 
or equal to 7,  
quarks are deconfined and chiral
symmetry is restored at zero temperature,
if the quark mass is lighter than a critical
value.
In this work we investigate 
the problem of what is
the condition on the number of flavors
for quark confinement and spontaneous chiral symmetry breaking
in the continuum limit \cite{lat96}.

As asymptotic freedom is the nature at short distance, one can apply
the perturbation theory to investigate the critical number for it.
However, because quark confinement and spontaneous chiral symmetry breaking
are due to non-perturbative effects, one has to apply a non-perturbative
method throughout the investigation of the critical numbers for them.
Therefore we employ lattice QCD for the study in this work,
since lattice QCD is such a theory constructed non-perturbatively. 

In order to investigate the condition on the number of flavors $N_F$
in the continuum limit,
we first clarify the phase structure at zero temperature 
for general number of flavors,
in particular, for $N_F \ge 7$. 
When the phase diagram becomes clear, 
we are able to see what kind of continuum limit exists
and eventually answer the question given above.

In Sec.~\ref{sec:latticeQCD} the formulation of lattice QCD is introduced.
The notations which will be used later,
and the Wilson quark action and the standard one-plaquette gauge 
action which we employ 
in this work are given. In Sec.~\ref{sec:chiral} quark mass is defined and
chiral property in the Wilson formalism of fermions is discussed.
Before going into details a brief summary is given Sec.~\ref{sec:summary}.
In Secs.\ref{sec:parameters} - \ref{sec:nf167}, 
details of numerical simulations
and results are given. Conclusions are given in Sec.~\ref{sec:conclusions}.

\section{Lattice QCD}
\label{sec:latticeQCD}
Lattice QCD is defined on a hypercubic lattice in 4-dimensional euclidean space
with lattice spacing $a$. 
A site is denoted by a vector $n=(n_1,n_2,n_3,n_4)$,
where $n_i$'s are integers.
A link with end points at the sites $n$ and $n+\hat\mu$
is specified by a pair $(n,\mu)$,
where $\hat\mu$ denotes a unit vector in the $\mu$ direction.

The gauge variable $U_{n,\mu}$ which is an element of SU(3) gauge group
is defined on the link $(n,\mu)$.
The action for gluons which we adopt is given by
\begin{equation}
S_{gauge}=\frac{1}{g^2} \sum_{n,\mu\neq\nu}{\rm Tr}
(U_{n,\mu}U_{n+\hat\mu,\nu}
U_{n+\hat\mu+\hat\nu,\mu}^{\dag} U_{n+\hat\nu,\nu}^{\dag}),
\label{eqn:gaction}
\end{equation}
where g is the gauge coupling constant.
This action is called the standard one-plaquette gauge action.
We usually use, instead of the bare gauge coupling constant $g$,
$\beta$ defined by
\begin{equation}
\beta=\frac{6}{g^2}.
\end{equation}

The quark variable is a Grassman number defined at a site.
It is well-known that a naive discretization of the Dirac action
\begin{equation}
S_{fermion}=\frac{a^3}{2}\sum_{n,\mu}(\bar{q}_{n}\gamma_{\mu}q_{n+\hat{\mu}}
-\bar{q}_{n+\hat{\mu}}\gamma_{\mu}q_{n})
+m_0a^4\sum_n \bar{q}_n q_n,
\label{eqn:navie}
\end{equation}
with $m_0$ being the bare fermion mass, 
leads to $16$ poles instead of one pole. This problem is called the species 
doubling. To avoid this problem K. Wilson proposed adding to the naive
discretized action a dimension 5 operator called the Wilson term
\begin{equation}
-a^3\sum_{n,\mu}(\bar{q}_{n}q_{n+\hat{\mu}}
-2\bar{q}_n q_n +\bar{q}_{n+\hat{\mu}}q_{n}),
\end{equation}
Rescaling the fermion field by
$q=\sqrt{2K}\psi$ and making the action gauge invariant, 
we obtain the Wilson fermion action
\begin{equation}
S_{Wilson}=a^3\sum_{n}[\bar{\psi}_n \psi_n
-K\{\bar{\psi}_{n}(1-\gamma_{\mu})U_{n,\mu}\psi_{n+\hat{\mu}}
+\bar{\psi}_{n+\hat{\mu}}(1+\gamma_{\mu})U_{n,\mu}^{\dag}\psi_{n}\}].
\end{equation}
where
\begin{equation}
K=\frac{1}{(m_0a+4r)},
\end{equation}
which is called the hopping parameter.

The full action $S$ is given by the sum of the gauge part
$S_{gauge}$ and the fermion part $S_{Wilson}$,
\begin{equation}
S=S_{gauge}+S_{Wilson}.
\end{equation}

The expectation value of an operator ${\cal O}(U,\psi,\bar \psi)$ is
given by
\begin{equation}
<{\cal O}>=\frac{1}{Z}\int \prod_{n,\mu}{\rm d}U_{n,\mu}
\prod_{n}{\rm d}\psi_n{\rm d}\bar{\psi}_n {\cal O}(U,\psi,\bar \psi)
{\rm exp}(S),
\end{equation}
where $Z$ is the partition function
\begin{equation}
Z=\int \prod_{n,\mu}{\rm d}U_{n,\mu}
\prod_{n}{\rm d}\psi_n{\rm d}\bar{\psi}_n
{\rm exp}(S),
\end{equation}
with ${\rm d} U_{n,\mu}$ being the Haar measure of SU(3).

In the case of degenerate $N_F$ flavors,
lattice QCD contains two parameters:
the gauge coupling constant $\beta =6/g^2$
and the hopping parameter $K$. In the non-degenerate case,
the number of the hopping parameters is $N_F$.

The lattice spacing $a$ is a function of the bare coupling constant
which is defined through the RG beta function.
The continuum limit is the limit where the lattice spacing $a$ 
goes to zero in such a way that physical quantities are
kept constant. In an asymptotically free case the limit corresponds
to the limit where the coupling constant $g$ vanishes.

At finite temperatures the linear extension in the time direction $N_t$
is much smaller that those in the spatial directions ($N_x, N_y, N_z$). 
The temperature $T$ is given by 
$$1/N_t a.$$
For gluons the periodic boundary condition is imposed, while
for quarks an anti-periodic boundary condition is imposed in the
time direction.

\section{Quark mass and chiral symmetry in the Wilson quark action}
\label{sec:chiral}
In the formalism of Wilson of quarks on the lattice, 
the flavor symmetry as well as C, P and T symmetries are exactly satisfied
on a lattice with a finite lattice spacing.
However, chiral symmetry is explicitly broken by the Wilson term
even for the vanishing bare quark mass $m_0=0$.
The lack of chiral symmetry causes much conceptual and technical
difficulties in numerical simulations and physics interpretation
of data. (See for more details Ref.~\cite{Stand26} and references cited there.)

The chiral property of the Wilson fermion action
was first systematically investigated
through Ward identities
by Bochicchio {\it et al.} \cite{Bo}.
We also independently proposed \cite{ItohNP} to define the current quark mass 
by
$$2 m_q <\,0\,|\,P\,|\,\pi\,>= - m_\pi <\,0\,|\,A_4\,|\,\pi\,>$$
where P is the pseudoscalar density and $A_4$ the fourth component of the
local axial vector current. We use this definition of the current quark
mass in this work. In general we need a multicative normalization factor for 
the axial current, which we have absorbed into the definition of the
quark mass, because this definition is sufficient for later use.
The value of the quark mass does not depend on
whether the system is in the high or the low temperature phase
when the gauge coupling constant is not so large: $\beta \ge 5.5$ for
$N_F=2$.

With this definition of quark mass,
the pion mass vanishes in the chiral limit where the quark mass vanishes
at zero temperature, for $N_F \le 6$.
However, in the deconfining phase at finite temperatures
the pion mass does not vanish in the chiral limit. 
It is almost equal to twice the lowest
Matsubara frequency $\pi/N_t$. This implies that the pion state is 
a free two-quark state. Quarks are not confined.
The pion mass is nearly equal to the scalar meson mass, and the rho meson mass
to the axial vector meson mass. The chiral symmetry is also manifest within
corrections due to finite lattice spacing.

Thus hadron masses in the chiral limit where the current quark mass 
vanishes clarifies the phase of the system at finite temperatures
for $N_F \le 6$.
We use the same strategy to discriminate the phase of QCD with various number
of flavors $N_F$: the pion mass at zero quark mass determines whether
chiral symmetry is spontaneously broken or not, and also whether quarks are
confined or not.
It should be noted that in QCD with $N_F$ flavors there is no known
order parameter for quark confinement. When the pion mass is about
twice the lowest Matsubara frequency we conclude that quarks are not confined.

\section{Brief Summary}
\label{sec:summary}
Before going into details let me give a brief summary, because
the following sections will be too technical for non-experts of lattice QCD.

\begin{figure}[tb]
\begin{center}
\leavevmode
\epsfxsize=5cm
\hspace{0.1cm}
\epsfbox{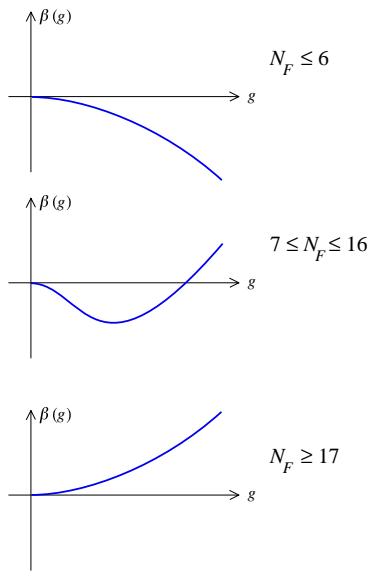}
\end{center}
\caption{Our conjecture for the beta fuction}
\label{conj_beta}
\end{figure}

\begin{figure}[tb]
\begin{center}
\leavevmode
\epsfxsize=5cm
\hspace{0.1cm}
\epsfbox{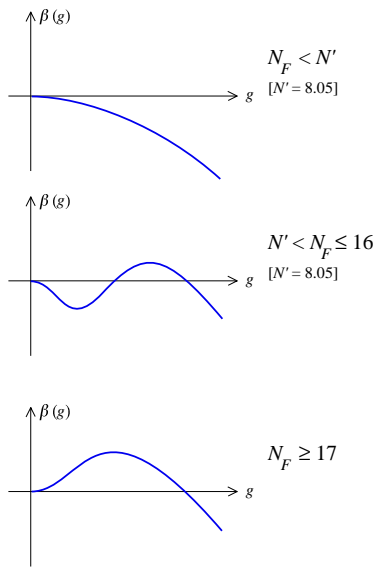}
\end{center}
\caption{Conjecture for the beta fuction by Banks and Zans}
\label{BanksZaks}
\end{figure}

In Fig.~\ref{conj_beta} is presented our conjecture in terms of the 
beta function 
of renormalization group.  Here $g$ is the bare gauge coupling constant.
When $N_F$ is equal or smaller than 6, 
the beta function is negative for all coupling
constants. On the other hand, when $N_F$ is equal or larger than 17,
it is positive. In the case when $N_F$ is between 7 and 16, it changes
sign from negative to positive with increasing $g$. Therefore there is
a non-trivial IR fixed point. The existence of such a fixed point was
first pointed out by Banks and Zaks. \cite{Banks1}
However, our picture is different those proposed by them.
Fig.~\ref{BanksZaks} is the conjecture made by Banks and Zaks \cite{Banks1}.
They assumed that
the beta function becomes negative for large enough coupling constant
for all $N_F$. This point is different from ours. 
Their critical value
is 9, which is also different from our value 7.

It is sometimes argued in the literature that quarks are confined in the
strong coupling constant. Although there is a proof for quark confinement
in the strong coupling constant in the pure $SU(3)$
gauge theory, there are no
such proofs in QCD with dynamical quarks. Our previous result \cite{previo} 
is a counter example for such an argument.

\begin{figure}[tb]
\begin{center}
\leavevmode
\epsfxsize=6.5cm
\epsfbox{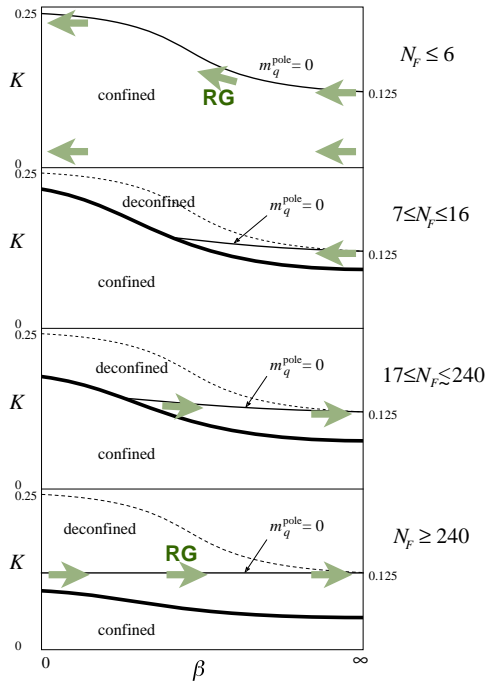}
\end{center}
\caption{Phase structure for various number of flavors 
$N_F$ and the RG flow.}
\label{RGflow}
\end{figure}

Fig.~\ref{RGflow} shows the phase structure in the $\beta$ - K plane for 
various number of flavors. When $N_F \le 6$, there is the chiral limit
$K_c(\beta)$, where the current quark $m_q$ vanishes, in the confining phase.
The value of $K_c(\beta)$
at $\beta = \infty$ is 1/8, which corresponds to the vanishing bare quark mass
$m_0=0$. As $\beta$ decreases, the $K_c$ increases, up to 1/4 at $\beta =0$.
If the action would be chiral symmetric, the chiral line should be a constant,
1/8.
The line $K=0$ corresponds to an infinite quark mass.
Quarks are confined for any value of the current quark mass for all values
of $\beta$.

\begin{figure}[tb]
\begin{center}
\leavevmode
\epsfxsize=6cm
\epsfbox{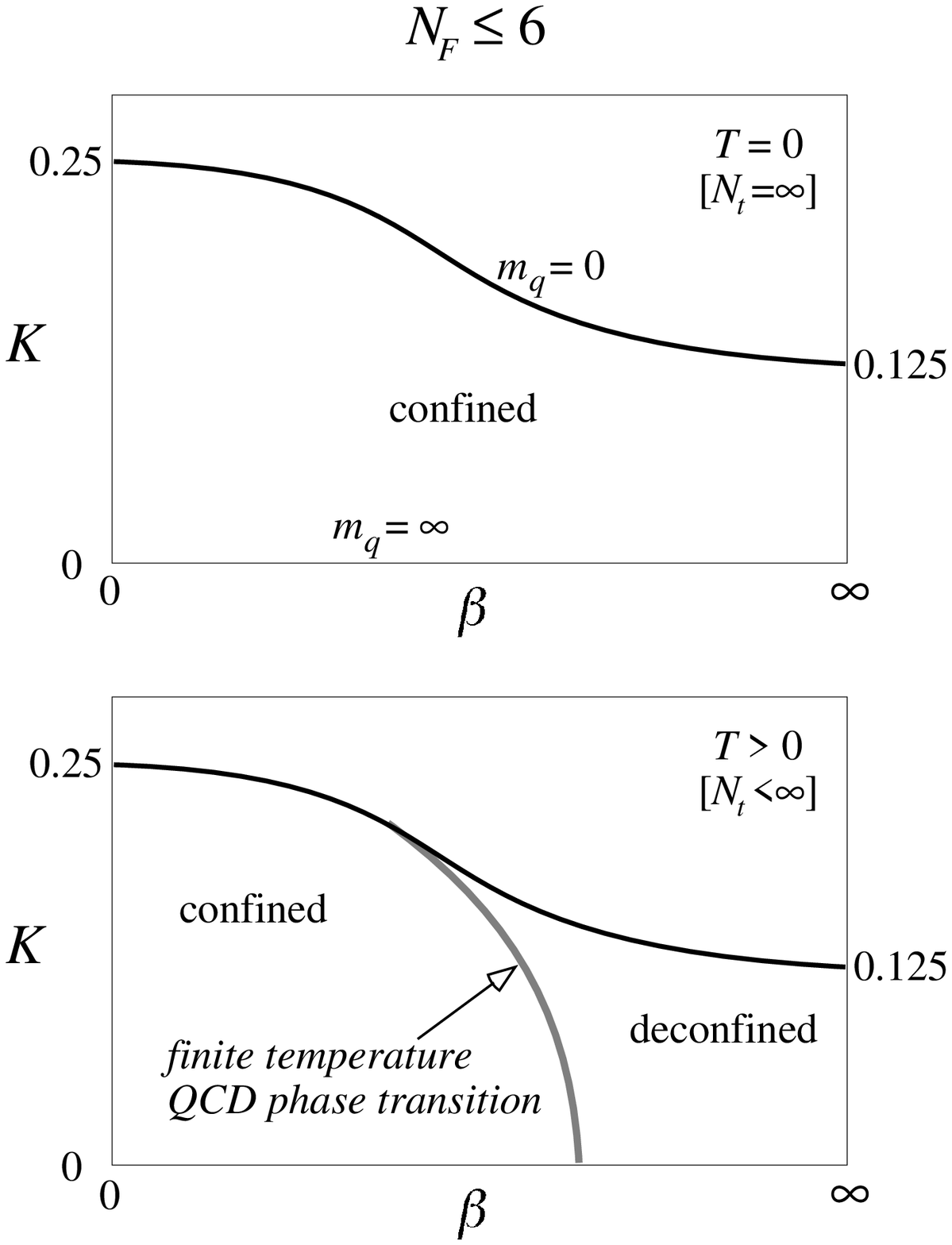}
\end{center}
\caption{The phase structure for $N_F \le 6$ at zero temperature
and at finite temperatures.}
\label{NF6_strc}
\end{figure}

When $N_F \ge 7$, there is no chiral limit in the confining phase.
There is a line of a first order phase transition from the confining phase
to a deconfining phase at a finite current quark mass for all values
of $\beta$. It is crucial that the point $K=1/8$ at $\beta=\infty$
does belong to a deconfining phase. The chiral line which starts from this
point hits the boundary between the confining phase and the deconfining phase
at finite $\beta$ as shown in Fig.~\ref{RGflow} for the case where $N_F$ 
is not too
large. This situation makes the interpretation of numerical data difficult;
the data around $\beta=4.5$ for $N_F=7$, 12 and 18 look very 
complicated.

In order to understand the structure of the deconfining phase more clearly, 
we increase $N_F$ up to 300,
because the region of the deconfining phase 
becomes larger with larger $N_F$,
and consequently the structure becomes simple,
as shown in Fig.~\ref{RGflow}.
When $N_F \ge 240$, the chiral line where the quark mass vanishes
goes straight from $\beta=\infty$ to $\beta=0$, without hitting
the phase transition boundary. We have made a MCRG study to investigate
the flow of RG along the massless line. The direction at $\beta=\infty$
is known from the perturbation theory, from left to right in the $\beta$ - K
plane. Our results of a MCRG study which will be described later implies that
the direction is identical all through the $\beta$ region. Thus there is
only a trivial IR fixed point at $\beta=\infty$. There are no other fixed
points where we are able to take a continuum limit of the theory with
quarks. Thus the theory is a trivial free theory.

As $N_F$ decreases, the massless quark line hits the boundary of a first
order phase transition. Except this point other general features are the 
same as those for $N_F \ge 240$. Massless quarks
exist only in the deconfining phase.
A continuum limit can be taken only at a fixed point on this line.
Although we have not investigated the flow of RG, we naturally assume
that the direction for $N_F \le 17$ is the same as that for $N_F \ge 240$. 
Thus the theory is also trivial for $N_F \le 17$.

When $7 \le N_F \le 16$, the phase structure seems to be not too much 
different from those for $N_F \ge 17$. However, the flow of RG at
$\beta=\infty$ is opposite to that for $N_F \ge 17$, because the theory is
asymptotically free. Therefore a continuum limit is governed by an
IR fixed point which should exist somewhere at finite value of $\beta$.
Unfortunately, as explained above, the massless line hits the boundary 
in the two parameter space. Therefore it is hard to see numerically
where the IR fixed
point exists. When one makes a RG transformation the trajectory moves into,
in general, a multi-parameter space. Thus within the two parameter space
we are investigating here it is difficult to pin down the position of the
IR fixed point. However, even without knowing the position we are 
able to discuss
the nature of the continuum limit. On the massless line around $\beta=4.5$,
the mass of the pion is roughly twice the lowest Matsubara frequency.
This implies that the quark is almost free. That is, an anomalous dimension
which is governed by the IR fixed point is small. 
The fact that the anomalous dimension is small suggests that
the IR fixed point exists at finite $\beta$:
If the IR fixed point would exist at $\beta=0$ or $g=\infty$, 
the anomalous dimension would be large or infinite. 
Anyway, the fact that
the mass of the pion is roughly twice the lowest Matsubara frequency
implies that quarks are not confined in this phase and the chiral symmetry
is not spontaneously broken. Thus this phase is a phase governed by
a non-trivial IR fixed point.

A salient fact for the above in this section is that
in the strong 
coupling limit, when the number of flavors $N_F$ is greater than 
or equal to 7,  
quarks are deconfined and chiral
symmetry is not spontaneously broken at zero temperature,
if the quark mass is lighter than a critical
value, which was shown in our previous work \cite{previo}.

In the following sections details of numerical simulations will be given.
One has to perform numerical simulations
on a lattice with a finite $N_t$. This implies that when $\beta$ becomes
large one encounters a finite temperature phase transition.
A schematic diagram is shown in Fig.~\ref{NF6_strc} for the case $N_F \le 6$.
In the case $N_F \ge 7$, one has to be careful about the existence of both
zero temperature transitions and finite temperature transitions.

\section{Simulation parameters}
\label{sec:parameters}

The lattice sizes are $8^2 \times 10 \times N_t$ 
($N_t =4$, 6 or 8), $16^2 \times 24 \times N_t$ ($N_t=16$)
and $18^2 \times 24 \times N_t$ ($N_t=18$).
We use an anti-periodic boundary condition for quarks 
in the $t$ direction and
periodic boundary conditions otherwise.
When the hadron spectrum is calculated, the lattice is duplicated 
in the direction of lattice size 10 for $N_t \le 8$, 
which we call the $z$ direction.
We call the pion screening mass simply the pion mass and 
similarly for the quark mass.

We use the hybrid R algorithm \cite{Ralgo} for the gauge configuration
generation.
As $N_F$ increases we have to decrease $\Delta\tau$, such as
$\Delta\tau$ =0.0025 for $N_F=240$, to reduce $O(\Delta\tau^2)$ errors. 
We have checked 
that the errors are sufficiently small selecting typical cases.

\begin{figure}[tb]
\begin{center}
\leavevmode
\epsfxsize=6cm
\epsfbox{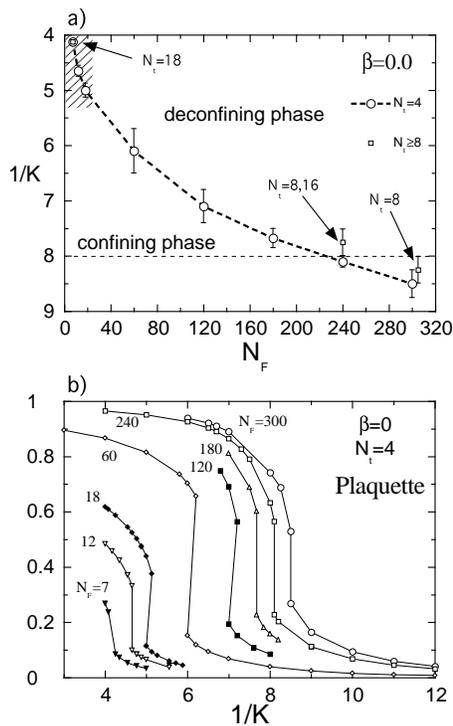}
\end{center}
\caption{a) The transition point $1/K_d$ at $\beta=0$ 
versus $N_F$ at $N_t=4$ and $N_t\ge 8$. 
For clarity, 
data at $N_t=8$ for $N_F=300$ is slightly shifted in the figure. 
Shaded region was investigated in our previous study 
\protect\cite{previo}.
b) Plaquette versus $1/K$ at $\beta=0$ for various number of $N_F$ at $N_t=4$. }
\label{kd.nf}
\end{figure}


\section{$K_d$ versus $N_F$}
We first study the case of $\beta=0$ as in the previous work \cite{previo}.
Fig.~\ref{kd.nf}a shows the results of the deconfining transition 
points $K_d$ at $\beta=0$ obtained on $N_t=4$ lattices 
for various numbers of $N_F$
 (see also Fig.~\ref{kd.nf}b).

The data of plaquette 
for $N_F=240$ at $\beta=0$ 
indicates that the deconfining transition is of first order 
and that the location of the transition point 
is independent of $N_t$ for large $N_t$ 
($1/K_d \simeq 8.1$ for $N_t=4$ and $1/K_d \simeq 7.75$ for $N_t=8$ 
and 16).
Therefore we conclude that the transition is bulk as is confirmed 
in our previous work for $N_F=7$ \cite{previo}.
In Fig.~\ref{kd.nf}a the transition points at $N_t\ge 8$ 
for $N_F=7$, 240 and 300 are also included.  
These values are roughly those for the bulk transition
points at zero temperature.
%

As $N_F$ increases up to 240, $1/K_d$ approaches toward or exceeds
the value 8, which is the value of $1/K$ for a massless free quark.
Because of this, for $N_F \simm{>} 240$, we are able to see 
a wide region of the deconfining phase 
in the whole range of $\beta$.
Therefore we
intensively investigate the case $N_F=240$, 
and then decrease $N_F$.

\section{$N_F=240$}

\begin{figure}[t]
\begin{center} 
\leavevmode
\epsfxsize=6cm
\epsfbox{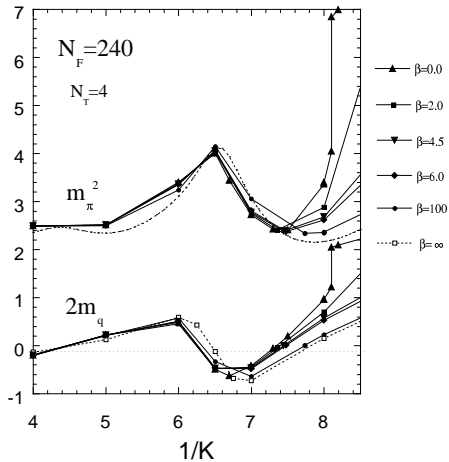}
\end{center}
\vspace{-0.5cm}
\caption{$m_\pi^2$ and 2$m_q$ versus $1/K$ for $N_F=240$ at $N_t=4$.}
\label{nf240.mass}
\end{figure}

Fig.~\ref{nf240.mass}
shows the results of $m_\pi^2$ and $2m_q$ for $N_F=240$ 
at $\beta=0$, 2.0, 4.5, 6.0, and 100 on the $N_t=4$ lattice.
A very striking fact is that the shape of $m_\pi^2$ and $2m_q$ 
as a function of $1/K$
only slightly changes for $1/K < 8$
when the value of $\beta$ decreases from $\infty$
down 0. 
(Note that the values at $\beta=\infty$ are those for free quarks.)
Only the position of the local minimum of $m_\pi^2$ at $1/K\simeq8$, 
which corresponds to the vanishing point of $m_q$, 
slightly shifts toward smaller $1/K$.
We obtain similar results also for $N_t=8$.

\begin{figure}[tb]
\begin{center} 
\leavevmode
\epsfxsize=7cm
\epsfbox{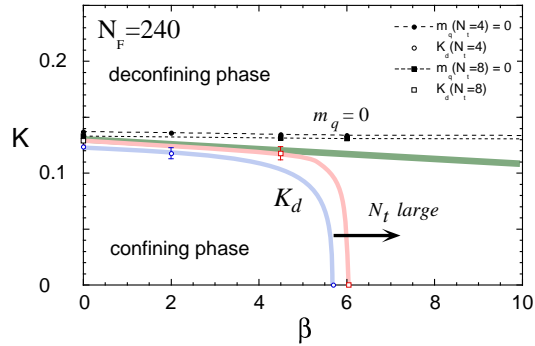}
\end{center}
\vspace{-0.5cm}
\caption{Phase diagram for $N_F=240$.}
\label{nf240.phase}
\end{figure}

From these data we propose the phase diagram
in Fig.~\ref{nf240.phase} for $N_F=240$.
The dark shaded line is the phase boundary between the 
confining phase and the deconfining phase at zero temperature.
When $N_t =4$ or 8, the boundary line bends down at finite $\beta$
as shown in Fig.~\ref{nf240.phase}, 
due to the finite temperature phase transition
of the confining phase.

The dashed line corresponds to the $m_q=0$ line. 
This line also corresponds to the minimum point of $m_\pi^2$.
We have also calculated the quark propagator in the Landau gauge and
checked that  chiral symmetry of the propagator,
$\gamma_5 G(z) \gamma_5 = - G(z)$,
is actually satisfied on the $m_q=0$ line. 

The results for the case $N_F=300$ are essentially the same as those
for $N_F=240$ except for very small shifts of the transition point
and the minimum point of $m_\pi^2$.

\section{Renormalization group flow}

The $m_q=0$ point at $\beta=\infty$ is the trivial IR
fixed point for $N_F \ge 17$. 
The phase diagram shown in Fig.~\ref{nf240.phase} suggests that 
there are no other fixed points on the $m_q=0$ line at finite $\beta$.

In order to confirm this, we investigate 
the direction of Renormalization Group (RG) flow along the $m_q=0$ line
for $N_F=240$, 
using a Monte Carlo Renormalization Group (MCRG) method.

One problem here is that it is practically impossible
to make simulations in the massless limit at zero temperature 
due to the existence of zero modes in the quark matrix.
Therefore, we impose an anti-periodic boundary condition 
in the $t$ direction.

We make a block transformation for a change of scale factor 2,
and estimate the quantity $\Delta \beta$ for the change of 
$a \rightarrow a'=2a$.
We generate configurations
on an $8^4$ lattice on the $m_q=0$ points at $\beta=0$ and 6.0
and make twice blockings. 
We also generate configurations
on a $4^4$ lattice 
and make once a blocking.
Then we calculate $\Delta\beta$ 
by matching the value of the plaquette at each step.\footnote{
It is known for the pure $SU(3)$ gauge theory, 
in particular in the deconfining
phase, that one has to make a more careful analysis 
using several types of Wilson loop
to extract a precise value of $\Delta \beta$.
We reserve elaboration of this point and a fine tuning of $1/K$ at each
$\beta$ for future works.
}

From the matching, we obtain $\Delta \beta \simeq 6.5$ for $\beta=0$ 
and $10.5$ for $\beta=6.0$.
The value obtained from the perturbation theory is 
$\Delta \beta \simeq 8.8$ at $\beta = 6.0$ for $N_F=240$.
The signs are the same and the magnitudes are comparable.
This suggests that the directions of RG flow on the $m_q=0$ line
at $\beta=0$ and 6.0 are the same as that at $\beta=\infty$ for $N_F=240$.
This further suggests that there are no fixed points at finite $\beta$.
All of the above imply that the theory is trivial in the case of $N_F=240$.

\section{$240 \ge N_F \ge 17$}


\begin{figure}[tb]
\begin{center} 
\leavevmode
\epsfxsize=7cm
\epsfbox{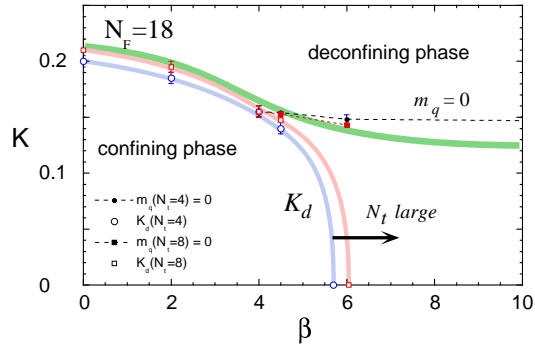}
\end{center}
\vspace{-0.5cm}
\caption{Phase diagram for $N_F=18$.}
\label{nf18.phase}
\end{figure}

Now we decrease the value of $N_F$ from 240.
%
When $\beta=6.0$, the shapes of $m_\pi^2$ are almost identical to each other,
except for a slight shift toward smaller $1/K$ as $N_F$ decreases.

When $\beta=0$, the boundary of the first order
phase transition between the deconfining phase and the confining phase
moves toward smaller $1/K$ and therefore the range of
the deconfining phase decreases: $1/K_d \simeq 8.5$, 8, 7.6, 7.2, 6.1 for
$N_F= 300$, 240, 180, 120 and 60, respectively, 
as shown in Fig.~\ref{kd.nf}.
For $N_F=18$, the value of $1/K_d$ decreases down to 5.0. 

Due to the fact that the confining
phase invades the deconfining phase, the massless line in the
deconfining phase hits the boundary at finite $\beta$ when $N_F$
becomes small. 
For example, in the case of $N_F=18$, 
as shown in Fig.~\ref{nf18.phase}, 
it hits at $\beta = 4.0$ --- 4.5.

Although the area of the deconfining phase decreases with 
decreasing $N_F$, 
the shape  and the position
of $m_\pi^2$ in the part of deconfining phase
only slightly change from $N_F=300$ to 18.
The values of $m_q$ as functions of $1/K$ 
also show only slight changes in the deconfining phase. 
These facts, combined with the perturbative result
that for $N_F \ge 17$, $\beta=\infty$ is the IR fixed
point, suggest that the RG flow along the massless quark in the
deconfining phase is the same as that for $N_F=240$.
In this case, the theory is trivial for $N_F \ge 17$.%
\footnote{
Although we have investigated only down to the $N_F=18$ case, 
we do not expect
any qualitative differences between the cases of $N_F=17$ and 18.
}

\section{$16 \ge N_F \ge 7$}
\label{sec:nf167}

\begin{figure}[tb]
\begin{center}
\leavevmode
\epsfxsize=7cm
\epsfbox{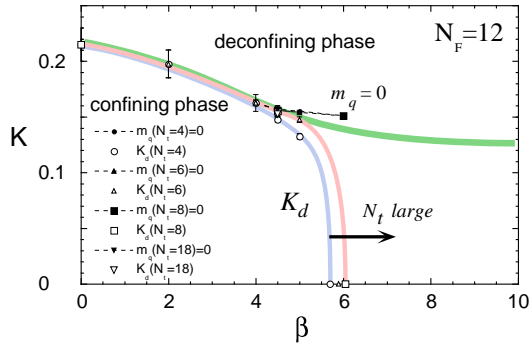}
\end{center}
\vspace{-0.5cm}
\caption{Phase diagram for $N_F=12$.}
\label{nf12.phase}
\end{figure}

When $N_F \le 16$, the theory is asymptotic free. 
On the other hand, 
quark confinement is lost for
$N_F \ge 7$ even in the strong coupling limit $\beta=0$ \cite{previo}.
Therefore the question is what happens 
for the cases $16 \ge N_F \ge 7$ in the continuum limit.

We have intensively simulated the cases $N_F=12$ and 7.
The phase diagram for $N_F=12$ is shown in Fig.~~\ref{nf12.phase}.
Although the gross 
feature of the phase diagram seems to be not too much 
different from the case of
$N_F=18$ which is shown in Fig.~\ref{nf18.phase},
the flow of RG at
$\beta=\infty$ is opposite to that for $N_F \ge 17$, because the theory is
asymptotically free. Therefore a continuum limit is governed by an
IR fixed point which should exist somewhere at finite value of $\beta$.
Unfortunately, as explained above, the massless line hits the boundary 
in the two parameter space. Therefore it is hard to see numerically
where the IR fixed
point exists. When one makes a RG transformation the trajectory moves into,
in general, a multi-parameter space. Thus within the two parameter space
we are investigating here it is difficult to pin down the position of the
IR fixed point. Even without knowing the position we are able to discuss
the nature of the continuum limit. On the massless line around $\beta=4.5$,
the mass of the pion is roughly twice the lowest Matsubara frequency.
This implies that the quark is almost free. That is an anomalous dimension
which is governed by the IR fixed point is small. 
The fact that the anomalous dimension is small suggests that
the IR fixed point exists at finite $\beta$:
If the IR fixed point would exist at $\beta=0$ or $g=\infty$, 
the anomalous dimension would be large or infinite. 
Anyway, the fact that
the mass of the pion is roughly twice the lowest Matsubara frequency
implies that quarks are not confined in this phase and the chiral symmetry
is not spontaneously broken. Thus this phase is a phase governed by
a non-trivial IR fixed point.
The theory in the continuum limit is a 
non-trivial theory with anomalous dimensions, 
however, without confinement. 
We reserve for future works
a detailed RG study to find a non-trivial fixed point in a wider
parameter space.

\section{Conclusions}
\label{sec:conclusions}
Based on numerical results combined with the result of 
the perturbation theory
we propose the following picture:
There are three categories depending on the number
of flavors $N_F$: a free theory for $N_F \ge 17$, a non-trivial theory
governed by a non-trivial IR
fixed point for $16 \ge N_F \ge 7$ and confinement and spontaneous
chiral symmetry breaking for $N_F \le 6$.

\section*{ACKNOWLEDGEMENTS}

I would like to thank K. Kanaya, S. Kaya, S. Sakai and T. Yoshie
for the collaboration of this work.

Numerical simulations are performed with 
HITAC S820/80 at KEK, and Fujitsu VPP500/30 and
QCDPAX at the University of Tsukuba.
This work is in part supported by 
the Grants-in-Aid of Ministry of Education,
Science and Culture (Nos.07NP0401, 07640375 and 07640376).

\end{document}